\documentstyle[preprint,aps,prbbib]{revtex}
\begin{document}
\draft

\title{ 
{\LARGE UNIVERSIT\'E DE GEN\`EVE}\\ 
{\scriptsize SCHOLA GENEVENSIS MDLIX}\\
\includegraphics{sigle.ps}
\rule[4mm]{1cm}{.5mm}\\
\vspace*{1cm}
Quantum-Dot Cascade Laser: 
Proposal for an Ultra-Low-Threshold Semiconductor Laser}
\author{Ned S. Wingreen$^1$ and C.~A.~Stafford$^{2}$}
\address{$\mbox{}^1$NEC Research Institute, 4 Independence Way, Princeton,
New Jersey 08540}
\address{$\mbox{}^2$D\'{e}partement de Physique 
Th\'{e}orique, Universit\'{e} de Gen\`{e}ve, 
CH-1211 Gen\`{e}ve 4, Switzerland}
\maketitle

\begin{center}
\fbox{UGVA-DPT 1996 / 05-926}\\
\end{center}

\begin{abstract}
We propose a quantum-dot version of the 
quantum-well cascade laser of Faist {\it et al.} 
[Science {\bf 264}, 553 (1994)]. The elimination 
of single phonon decays by the three-dimensional
confinement implies a several-order-of-magnitude
reduction in the threshold current. The requirements
on dot size (10-20nm) and on dot density and uniformity 
[one coupled pair of dots per (180nm)$^3$ with 
5\% nonuniformity] are close to current technology. 
\end{abstract}

\newpage
\tighten
\narrowtext

The recent demonstration by Faist {\it et al.}\cite{faist} of
a laser based on a cascade of coupled quantum wells has opened up new
possibilities in semiconductor lasers. Here we explore one
possibility aimed at reducing the threshold current: a version
of the Faist {\it et al.} laser based on quantum dots rather than quantum wells.
While there have been various proposals for low-threshold quantum-dot 
lasers,\cite{ehdots} all have been based on electron-hole recombination. 
Attempts to realize a quantum-dot laser of the electron-hole
type have been unsuccessful, most likely due to slow energy 
relaxation of electrons in the dots, leading to poor recombination 
efficiency.\cite{benistry}

The quantum-dot cascade laser we propose here offers the advantages 
of an intrinsically strong and narrow gain spectrum, with a minimal rate of
nonradiative decays.  As in the quantum-well cascade laser, the current
directly pumps the upper lasing level so there is no problem
of slow relaxation.\cite{benistry} However, unlike the quantum-well cascade
laser, nonradiative decay by phonon emission can be eliminated.
Since the nonradiative 
rate of decay due to phonon emission in quantum wells is 3000 times the
radiative decay rate,\cite{faist} elimination of phonon decays is a 
priority.  To eliminate phonon emission in the proposed quantum-dot
scheme requires dots smaller than  10-20 ${\rm nm}$
in all three dimensions. This is the primary technological difficulty,
but there is reason to believe that such dimensions can be 
achieved.\cite{dotsizea,dotsizeb}

In what follows, we will describe the proposed quantum-dot laser
in more detail and compare it to the 
quantum-well cascade laser.\cite{faist} The dot size requirements 
will be estimated as well as the dot density and uniformity requirements. 
The latter follow from a comparison of the gain coefficient 
to the typical effective loss in semiconductor injection
lasers. Finally, the threshold current for lasing is estimated
from the total rate of spontaneous emission.

It is profitable to compare and contrast the proposed
quantum-dot laser and the quantum-well 
laser of Faist {\it et al.}\cite{faist}
using the simplified conduction-band 
energy diagram in Fig. 1, which suffices for both.
The diagram shows two electronically coupled 
dots or wells.\cite{threewells} 
Photons are generated 
by the transition of an electron from the first excited state 
to the ground state of the coupled dots or coupled wells.
In both cases, electrons are injected directly into the excited 
state by a current tunneling through the upstream barrier. 
Once an electron is de-excited it escapes 
quickly through the downstream barrier, so that photon absorption 
is negligible.

The essential difference between the quantum-dot and quantum-well 
lasers is that in the former the excited and ground electronic 
levels shown in Fig. 1 represent truly discrete states, while in 
the latter each represents the bottom of a continuous band of states.
Specifically, in the quantum-well case electrons form bands due
to their free movement in the two dimensions transverse 
to the direction of conduction-band energy variation shown in Fig. 1
({\it i.e.}, the  direction of current flow). 
As a consequence, the 
dominant electronic decay
mechanism in the quantum wells is nonradiative,
involving emission of an optic phonon rather than a photon.
Since the bands are continuous in energy such transitions
are always allowed, and since the electron-optic-phonon 
coupling is much stronger than the electron-photon coupling
such nonradiative transitions will always dominate the
radiative ones. 

In contrast, in the quantum-dot laser the rate of radiative 
decay may dominate the nonradiative rate.
Since the excited and ground states of the coupled dots are
discrete levels, nonradiative decays will involve emission of
a phonon at the difference energy. In general, phonon energies 
form a continuous band so that such one phonon decays are 
allowed. However, if the difference energy is larger than
the largest phonon energy ({\it e.g.}, the optic phonon energy
at  $\hbar \omega_{LO} = 36\, {\rm meV}$ in GaAs),
then no single phonon can carry away all 
the electronic 
energy. Multiphonon decay processes are still allowed but the 
rate of these is negligible (except in certain  narrow energy 
bands\cite{inoshita}). The dominant decay
mechanism in dots can therefore be photon emission with a consequent
enhancement of overall efficiency.  

The size of each of the coupled dots is strongly constrained by 
the requirement that the energy difference between the excited
and ground states exceeds the optic-phonon energy $\hbar \omega_{LO}$. 
Specifically, the energy difference between the lowest states of 
one of the dots in isolation must exceed $\hbar \omega_{LO}$. The resulting 
maximum dot size $L$ can be estimated from the energy spacing in 
a square well of size $L$,
\begin{equation}
 \frac{3 \pi^2 \hbar^2}{2 m^* L^2} >
 \hbar \omega_{LO}.
\label{dotsize}
\end{equation}
For GaAs, with an effective mass $m^* = 0.067 m$, this implies
dots smaller than $L \simeq 20{\rm nm}$ in all three dimensions.
Figure 2(a) shows a schematic array of pairs of such coupled dots sandwiched
between conducting sheets. The necessary size scales are close to 
current technology:
dot arrays involving single
quantum dots have been fabricated by electron-beam lithography with dot diameters 
of 57 nm,\cite{dotsizea}
and arrays with dot diameters of 25 nm have 
been achieved via self-assembled growth.\cite{dotsizeb}

More is required than just small dots, however, since a laser
also requires gain. Laser action will only occur if the 
gain coefficient $\gamma(\omega)$ exceeds the distributed loss, 
\begin{equation}
 \gamma(\omega) > \alpha_I + \alpha_M, 
\label{threshold}
\end{equation}
where $\alpha_I$ is the bulk loss and $\alpha_M = (1/L) \log(1/R)$
is the loss through the mirrors. 
Relation (\ref{threshold}) jointly constrains the 
minimum density of dot pairs and the uniformity of dot sizes.
The gain is proportional to the three-dimensional density of
coupled dots $N$,\cite{saleh}
\begin{equation}
 \gamma(\omega) = f N \sigma(\omega),
\label{gain}
\end{equation}
where $f$ is the fraction of coupled dots with an electron in the excited 
state (we neglect the small fraction of dots with an electron in the
ground state) and $\sigma(\omega)$ is the cross section.
It is convenient to write $\sigma(\omega)$ as the product of an 
oscillator strength $S$ and a normalized lineshape function $g(\omega)$,
\begin{equation}
 \sigma(\omega) = S g(\omega).
\label{xsection}
\end{equation}
In the dipole approximation, the oscillator strength is given 
by\cite{merzbacher}
\begin{eqnarray}
S &=& \frac{4 \pi^2 \alpha\, \omega_{fi}}{n} \ 
         | \langle f | {\bf r} \cdot \hat {\bf e} | i \rangle|^2 \nonumber \\
 &\simeq& \frac{4 \pi^2 \alpha\, \omega_{fi}}{n} 
         \left(\frac{t d}{\hbar \omega_{fi} } \right)^2,
\label{Seq}
\end{eqnarray}
where $\alpha = e^2/\hbar c \simeq 1/137$ is the fine structure constant,
$n$ is the index of refraction, and 
$\omega_{fi}$ is the transition frequency between initial
and final states.
The dipole matrix element between initial and final states
$  \langle f | {\bf r} \cdot \hat {\bf e} | i \rangle$ 
projects the polarization direction $\hat {\bf e}$
on the dipole moment. In the coupled dot, the transition dipole moment 
lies purely along the current direction so the radiation will
be polarized in that direction. In the second line of (\ref{Seq}), 
we have approximated the dipole matrix element by the product of the
distance between the dots and interdot hybridization 
$t/\hbar \omega_{fi}$, where $t$ is the tunnel coupling between
dots.\cite{hybrid}
The remaining factor in the cross section is the normalized 
lineshape function $g(\omega)$. It is realistic to assume that
inhomogeneous broadening due to disorder will determine the
lineshape. Taking, for convenience, a Lorentzian lineshape with FWHM
disorder broadening of $\Delta \omega$, one finds a peak gain
coefficient of\cite{faistpre}
\begin{equation}
 \gamma_{\rm peak} = \frac{2f N S}{\pi \Delta \omega}.
\label{peakgain}
\end{equation}
By equating the peak gain in (\ref{peakgain}) to the total loss,
we can state the joint requirement on density and uniformity for a 
functional quantum-dot laser. The distributed loss for a 
semiconductor injection laser is at least 10 cm$^{-1}$.\cite{saleh}
The interdot hybridization $t/\hbar \omega_{fi}$
must be sufficiently  small that 
the spontaneous emission rate $w_{\rm sp}$ dominates the leakage
rate from the excited state through the downstream barrier.
In turn, the spontaneous emission rate must be smaller than
the escape rate from the ground state.
Assuming a fixed escape rate
$\Gamma$ through the downstream barrier, these inequalities imply 
\begin{equation}
 (t/\hbar \omega_{fi})^2\,  \Gamma < w_{\rm sp} < \Gamma,
\label{ineqs}
\end{equation}
which clearly limits the hybridization to 
$(t/\hbar \omega_{fi})^2 \stackrel{<}{\scriptscriptstyle \sim} 1/10$. 
(However, this condition can be relaxed by additional
bandstructure engineering.\cite{threewells})
Further, assuming a transition energy of 100 meV, interdot spacing 
of $d = 10 {\rm nm}$, and index of refraction $n=3$,
we find an excited coupled-dot density to broadening energy ratio of 
\begin{equation}
\frac{fN}{\hbar  \Delta \omega} \simeq 
 1.6 \times 10^{16} {\rm cm^{-3} eV^{-1}}.
\label{fom}
\end{equation}
Hence a 10\% disorder broadening of the transition energy 
(5\% nonuniformity), and an excited fraction
near $f=1$, implies a minimum density of one coupled
dot pair per $(180{\rm nm})^3$ volume.

To achieve this average density of coupled dots throughout
the region occupied by the lasing mode requires
a true three-dimensional structure. One can envision 
a layered structure, each layer consisting of a dense array
of coupled quantum dots, as sketched in Fig. 2(b), with an overall
density satisfying the conditions for gain. The stacking
of arrays of quantum dots is analogous to the cascade 
of coupled quantum wells employed by Faist {\it et al.},\cite{faist} so
the resulting device should properly be called a 
``quantum-dot cascade laser". 

Finally, we can estimate the threshold current for such a
device. Since radiative decays dominate, the current
flowing through each pair of dots need only be adequate
to replenish losses due to spontaneous emission.
The total spontaneous emission rate is\cite{merzbacher}
\begin{eqnarray}
w_{\rm sp}  &=& \frac{ 4\, \alpha\, n\, \omega_{fi}^3}{3 c^2} 
      | \langle f | {\bf r}| i \rangle|^2 \nonumber \\
 &\simeq& \frac{4\, \alpha\, n\, \omega_{fi}}{3}
         \left(\frac{t d}{\hbar c } \right)^2.
\label{wsp}
\end{eqnarray}
Using the same parameters as above one finds a threshold
current of $J_{t} = e w_{\rm sp} \simeq 1.6 {\rm pA}$
per coupled dot pair. For a uniform array of dots in 
three dimensions this gives a current density of 
$4.9 {\rm mA/cm^2}$. While it is not fair to compare
the calculated performance of a proposed device to the
actual performance of a real device, it is still striking
that this threshold current density is some 
six-and-a-half 
orders of magnitude lower than the quantum-well cascade
laser value of $14 {\rm kA/cm^2}$.\cite{faist}

In addition to the low threshold current, the quantum-dot
cascade laser offers several other advantages. The operation
should be essentially temperature independent provided $kT$ is 
smaller than the level spacing in the dots. Since the levels
are discrete there is no thermal broadening, and since phonon
decay processes are eliminated there is no increase in the 
decay rate with increasing phonon occupation. Finally, the
operation frequency is in principle tunable by the applied
bias as in the quantum-well structure.\cite{faistapl}

In conclusion, we have proposed and analyzed a version of
the quantum cascade laser\cite{faist} based on quantum
dots rather than quantum wells. The quantum dot version
offers the possibility of a several-order-of-magnitude 
reduction in the threshold current by eliminating 
single phonon decays. The constraints on dot size (10-20{\rm nm}),
and dot density and uniformity [one coupled dot pair
per (180{\rm nm})$^3$ with 5\% nonuniformity] are close to 
current technology. We hope to have stimulated interest
in constructing a low-threshold semiconductor laser based on 
electronic transitions in quantum dots.

We thank Federico Capasso for providing us with unpublished
results 
and Judy Sonnenberg for technical assistance with the 
manuscript.  
One of us (C.\ A.\ S.) acknowledges support from the 
Swiss National Science Foundation.

\begin{figure}
\vbox to 16cm {\vss\hbox to 17cm
  {\hss\
    {\includegraphics{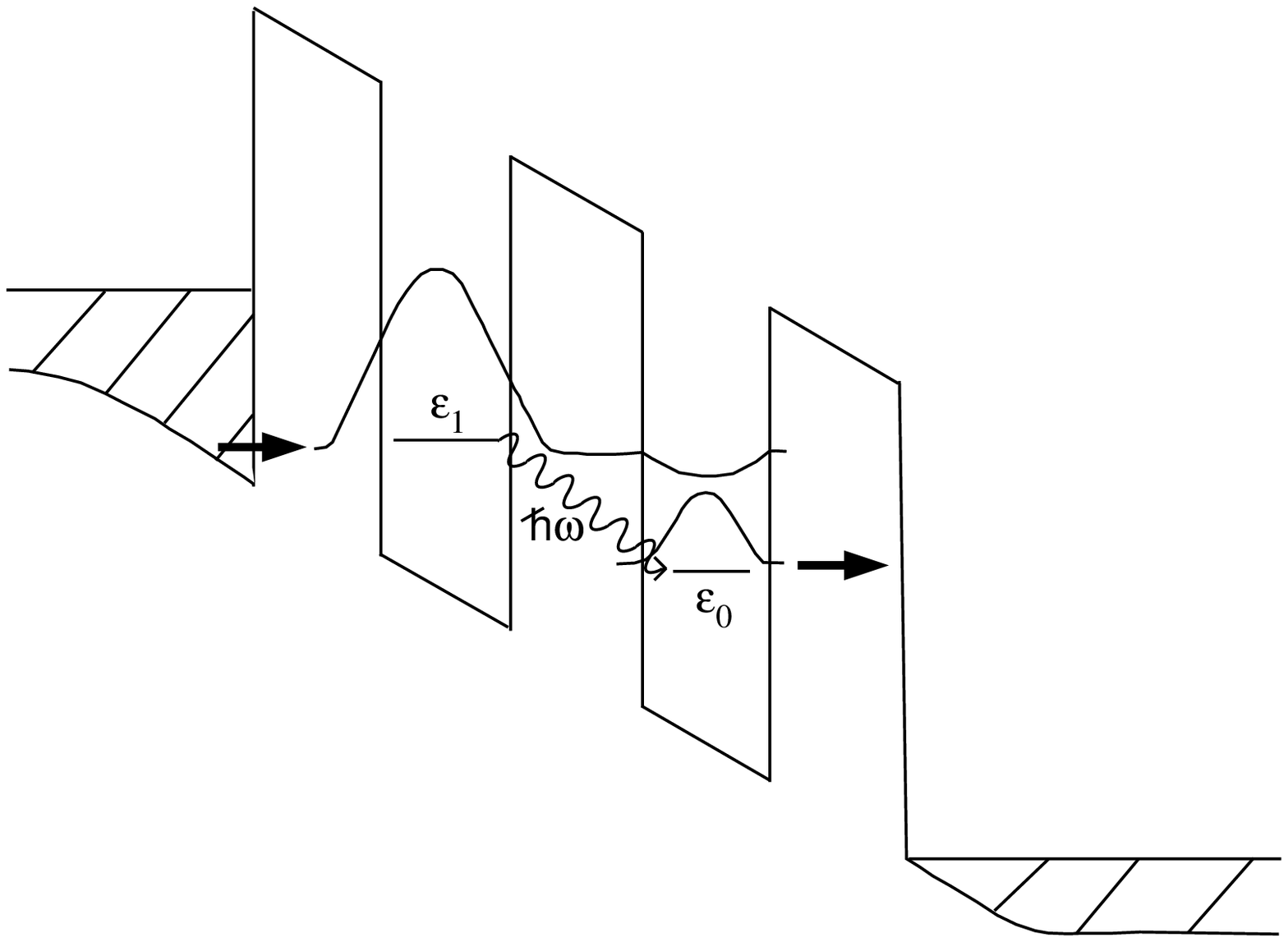}}
   \hss}
}
\caption{Schematic conduction-band energy diagram of active region of
proposed quantum-dot cascade laser. For low-threshold-current
operation the energy difference between the first excited state
and the ground state of the coupled  dots, $\epsilon_1 - \epsilon_0$, must
be larger than all phonon energies.}
\label{fig1}
\end{figure}

\begin{figure}
\vbox to 10cm {\vss\hbox to 17cm
  {\hss\
    {\includegraphics{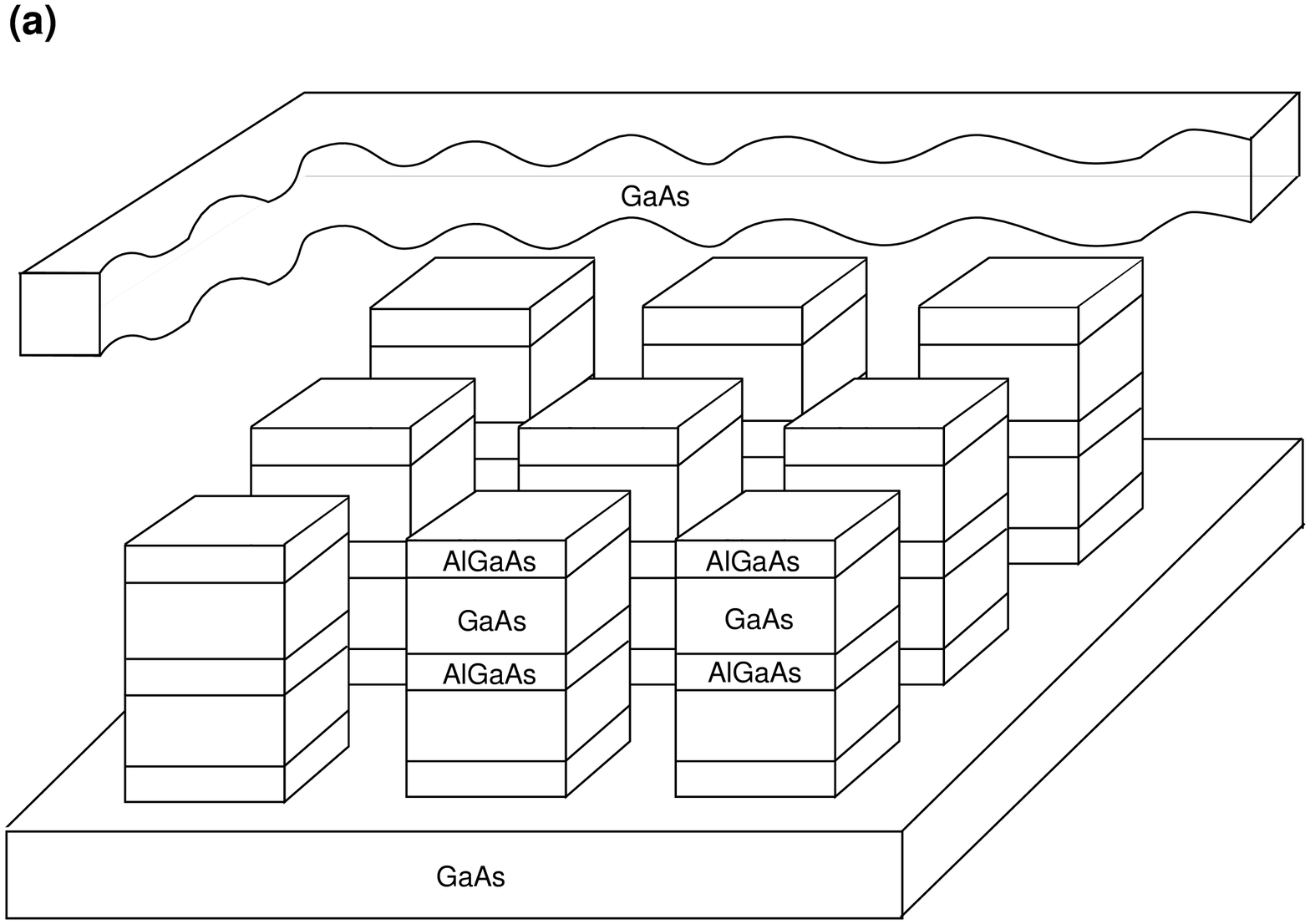}}
   \hss}
}
\vbox to 10cm {\vss\hbox to 17cm
  {\hss\
    {\includegraphics{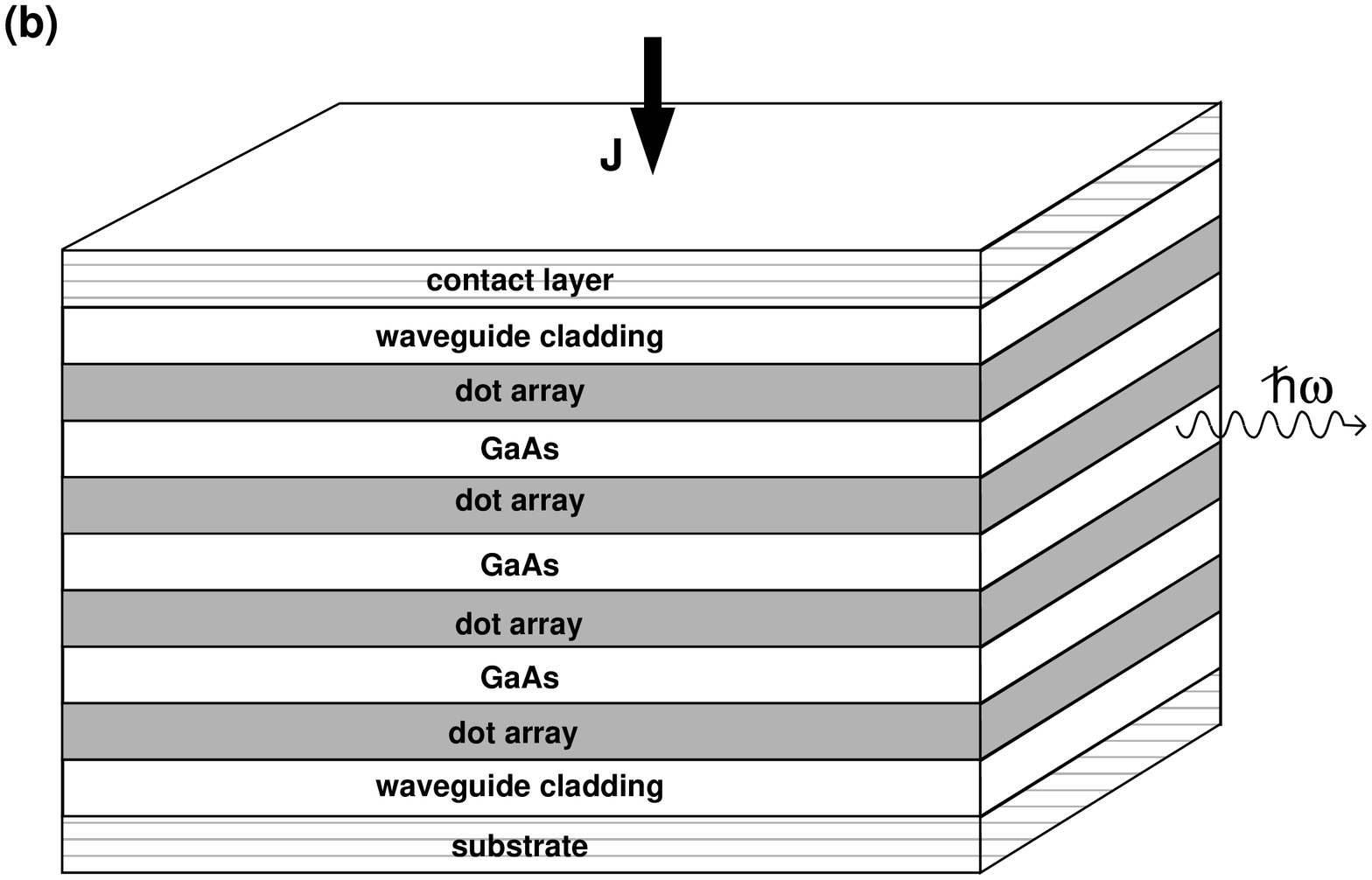}}
   \hss}
}
\caption{(a) Schematic representation of coupled-quantum-dot array.
For laser operation the space between pairs of dots 
(shown as pillars) must be insulating so that a vertically directed 
current is constrained to flow through the dots. The conducting
regions immediately above and below each pair of dots may be
connected to form a continuous sheet as shown. (b) Stacked layers
of coupled-quantum-dot arrays, in the cascade configuration
developed for the quantum-well laser by Faist {\it et al.} [Ref. 1].}
\label{fig2}
\end{figure}


\begin{references}

\bibitem{faist} J. Faist, F. Capasso, D. L. Sivco, C. Sirtori, A. L. 
Hutchinson, and A. Y. Cho,  {Science} {\bf 264}, 553 (1994).

\bibitem{ehdots} Y. Arakawa and H. Sakaki, {Appl. Phys. Lett}
{\bf 40}, 939 (1982); S. Schmitt-Rink, D. A. B. Miller, and D.S.
Chemla, {Phys. Rev. B} {\bf 35}, 8113 (1987);
M. Yamanishi and Y. Yamamoto, {Jpn. J. Appl. Phys.}
{\bf 30}, L60 (1991).

\bibitem{benistry} H. Benistry, C. M. Sotomayor-Torr\`es, and 
C. Weisbuch, {Phys. Rev. B} {\bf 44}, 10945 (1991).

\bibitem{dotsizea} T. D.  Bestwick, M. D. Dawson,  A. H. Kean, and G. Duggan,
    {Appl. Phys. Lett.} {\bf 66}, 1382 (1995). 

\bibitem{dotsizeb} 
    D. Leonard, M. Krishnamurthy, C. M. Reaves, S. P. Denbaars, and 
    P. M. Petroff,
    {Appl. Phys. Lett.} {\bf 63}, 3203 (1993); K. Nishi, R. Mirin,
    D. Leonard, G. Medeiros-Ribeiro, P. M. Petroff, and A. C. Gossard, 
    submitted to {J. Appl. Phys.} 

\bibitem{threewells}
In practice, quantum-well cascade 
lasers contain three wells and/or a superlattice Bragg reflector 
to control the rate of tunneling escape [J. Faist, F. Capasso,
C. Sirtori, D. L. Sivco, A. L. Hutchinson, and A. Y. Cho, 
{Appl. Phys. Lett.}  {\bf 66}, 538 (1995)]. 

\bibitem{inoshita} The multiphonon rate can be significant in a
narrow band around the optic phonon energy [T. Inoshita and H. Sakaki, 
Solid-State Electronics
{\bf 37}, 1175 (1994)]. The rate can also be significant 
at low multiples of the optic-phonon energy, with the total rate falling
off as $(0.04)^N$ for $N$-optic-photon emission in GaAs. However, these
resonances can be avoided by proper tuning of the energy
difference between the excited and ground states.





\bibitem{saleh} B. E. A. Saleh and M. C. Teich, {\em Fundamentals
of Photonics}, Wiley, New York, 1991.

\bibitem{merzbacher} See, {\it e.g.},  E. Merzbacher, {\em Quantum
Mechanics}, Wiley, New York, 1961, 2nd ed. 1991.

\bibitem{hybrid} The interdot hybridization $t/\hbar \omega_{fi}$
gives the ratio of amplitudes in the dots, {\it i.e.} a hybridization
of $\sqrt{1/10}$ implies a probability of $(\sqrt{1/10})^2 = 1/10$ of finding 
the first excited state electron in the downstream dot. 

\bibitem{faistpre} An equivalent expression is employed 
to analyze the performance of the quantum-well cascade laser
[J. Faist, F. Capasso,
D. L. Sivco, A. L. Hutchinson, C. Sirtori, S. N. G. Chu, and A. Y. Cho, 
  {Appl. Phys. Lett.} {\bf 65}, 2901 (1994)].

\bibitem{faistapl} J. Faist, F. Capasso,
C. Sirtori, D. Sivco, A. L. Hutchinson, S. N. G. Chu, and A. Y. Cho, 
 {Appl. Phys. Lett.} {\bf 64}, 1144 (1994).

\end{references}
\end{document}